\documentclass[10pt,a4paper]{article}
\usepackage{graphicx}

\title{Casimir force between designed materials:\\
what is possible and what not}
\author{C. Henkel$^1$ and K. Joulain$^2$
\\
$^1$ Institut f\"{u}r Physik, Universit\"{a}t Potsdam,
Germany \\
$^2$ Laboratoire d'Etudes Thermiques,
Ecole Nationale Sup\'erieure
\\
de M\'ecanique A\'eronautique, Poitiers, France}
\date{20 October 2005}

\begin{document}

\maketitle

\begin{abstract}
    We establish strict upper limits for the Casimir interaction
    between multilayered structures of arbitrary dielectric or diamagnetic
    materials. We discuss the appearance of different power laws
due to frequency-dependent material constants. Simple analytical
expressions are in good agreement with numerical calculations based on
Lifshitz theory. We discuss the improvements required for current
(meta) materials to achieve a repulsive Casimir force.
\\
Dated: 20 Oct 2005, \emph{Europhysics Letters}, in press.
\end{abstract}

\section{Introduction}

The optical properties of materials that show both a
dielectric and magnetic response, have recently attracted
much attention (see \cite{Ramakrishna05} for a review). A number
of striking phenomena like perfect lensing and a reversed Doppler
effect have been predicted, and experimenters have
begun to explore the large parameter space of structural units
that can be assembled into artificial materials.
Breakthroughs have been reported on 
the way towards designed susceptibitilies in the near-infrared and
visible spectral range \cite{Pendry04b,Wegener04}.
Quantum electrodynamics in meta materials has recently been explored 
with particular emphasis on left-handed or negative-index materials 
\cite{Klimov02d,Fleischhauer05a}. We discuss here to what
extent the Casimir interaction between two meta material plates
can be manipulated by engineering their magneto-dielectric response.
Strict limits for the Casimir interaction are proven
that apply to all causal
and linear materials, including both bulk and multilayer structures.
We illustrate these results by computations of the
Casimir pressure, considering materials with
frequency-dispersive response functions like those encountered
in effective medium theories. We derive power law exponents
and prefactors and find that a strongly modified Casimir 
interaction is possible
in a range of distances around the resonance wavelength of
the response functions. We give estimates for
the required temperature range and structure size: it is
not unreasonable to expect that improvements in fabrication
and detection will allow for experimental observations.

One of the most striking changes to the Casimir interaction is a cross
over to repulsion. This has been predicted previously for idealized
magnetodielectric materials \cite{Boyer74,Kenneth02,Boyer03} and
objects suspended in a liquid \cite{Hartmann91,Israelachvili}.
In the latter case, repulsion has been observed experimentally with
colloidal particles \cite{Sigmund01} and is also used in a recent
proposal for measuring Casimir torques \cite{Capasso05}. Casimir
repulsion between mirrors separated by vacuum requires a strong magnetic
response \cite{Kenneth02,Tomas05} that hardly occurs in conventional
ferromagnets \cite{Camley98,Kenneth02c-2}. Indeed, to manipulate the
Casimir force in the micrometer range and below, where it can be
conveniently measured, the key challenge is to achieve a magnetic
susceptibility at high frequencies, approaching the visible range.
Now, there is a well-known argument due to Landau, Lifshitz, and
Pitaevskii that $\mu( \omega ) = 1$ in the
visible \cite{Landau10}. This objection, however, only applies to
materials whose magnetization is of atomic origin, where the magnetic
susceptibility is $\chi_{\rm m} \sim (v/c)^2 \ll 1$. An array of split
ring resonators with sub-wavelength size typically gives, on the
contrary, $\chi_{\rm m} \sim (\omega/\vartheta)^2 f \sim 1$, where
$\vartheta$ is the resonance frequency and $f$ the filling factor
\cite{Pendry99d,Smith04a}. As we illustrate below, artificial materials 
that are structured on the sub-micron scale are promising candidates
for a strongly modified Casimir interaction.

\section{Lifshitz theory}

For two perfectly conducting plates
held at zero temperature and separated by a distance
$d$, Casimir derived a force per
unit area given by
\(
F_{C} = 
\pi^2 \hbar c /( 240 \, d^4 )
\)
\cite{Casimir48b}.
We use the convention that $F_{C} > 0$
corresponds to attraction.
For linear media with complex, frequency-dependent material
parameters, the force can be computed from Lifshitz theory 
\cite{Lifshitz56}. This expression has been re-derived,
for plates of arbitrary material and for multilayer mirrors, 
using
different methods~\cite{Parsegian69,Parsegian70b,Ninham71,Schram73,%
Schwinger78,%
Kupiszewska92,Spruch95,Klimchitskaya00,Villarreal01a,Tomas02,Genet03c}.
At finite temperature, it can be written in the form
\begin{eqnarray}
    F_{L} &=& 2 k_{B} T
    {\kern 4.5ex\raisebox{0.25ex}{$'$}\kern -4.5ex}\sum\limits_{n=0}^{\infty} 
    \int\limits_{\xi_{n}/c}^{\infty} \! \frac{ {\rm d}\kappa }{ 2 \pi } \,
    \kappa^2 
    \sum_{\lambda}
    \left( 
    \frac{ {\rm e}^{2 \kappa d } }{ 
    r_{\lambda 1}
    r_{\lambda 2} 
    } - 1
    \right)^{-1}\!\!
    ,
    \label{eq:Lifshitz-formula}
\end{eqnarray}
where the sum is over the imaginary Matsubara frequencies
$\omega_{n} = {\rm i} \xi_{n} \equiv 2\pi {\rm i} n k_{B} T / \hbar$
(the $n=0$ term being multiplied by $1/2$),
and $\kappa$ is related to the wave vector component perpendicular 
to the mirrors, 
$k_{z} = (\omega^2_{n}/c^2 - k_{x}^2 - k_{y}^2)^{1/2} 
\equiv {\rm i}\, \kappa$.
The $r_{\lambda \alpha}$ ($\lambda = {\rm TE}, \, {\rm TM}$, 
$\alpha = 1, \, 2$) are the reflection coefficients at mirror $\alpha$ 
for electromagnetic waves with polarization $\lambda$ 
\cite{Parsegian70b,Ninham71}. 
For homogeneous, thick plates, they are given by 
\begin{eqnarray}
r_{\rm TM}
    &=&
    \frac{ \varepsilon ( {\rm i} \xi_{n} ) c \kappa - 
      \sqrt{ \xi_{n}^2 (\varepsilon ( {\rm i} \xi_{n} ) \mu (
      {\rm i} \xi_{n} ) - 1) + \kappa^2 c^2 
      } 
    }{ 
    \varepsilon ( {\rm i} \xi_{n} ) c \kappa + 
      \sqrt{ \xi_{n}^2 (\varepsilon ( {\rm i} \xi_{n} ) \mu (
      {\rm i} \xi_{n} ) - 1) + \kappa^2 c^2 }
    }
    \label{eq:Fresnel-r}
\end{eqnarray}
(exchange $\varepsilon$ and $\mu$ for $r_{\rm TE}$).
The zeros of 
$D_\lambda \equiv {\rm e}^{2 \kappa d } / ( 
    r_{\lambda 1}
    r_{\lambda 2} 
    ) - 1$
at real frequencies define the eigenmodes of the 
cavity formed by the two mirrors. 

\section{Strict limits}

To derive upper and lower limits for $F_{L}$, we use that
the Kramers-Kronig relations~\cite{Landau10} imply real and positive 
material functions at imaginary frequencies,
$\varepsilon( {\rm i}\xi ) \ge 1$,
provided the material is passive (non-negative absorption 
${\rm Im}\,\varepsilon( \omega ) \ge 0$). As a consequence, the 
Fresnel formulas~(\ref{eq:Fresnel-r}) imply $-1 \le r_{\lambda \alpha} \le 1$,
and we find 
\begin{equation}
- \frac{ 1 }{ {\rm e}^{2\kappa d} + 1 } \le
\frac{ 1 }{ D_\lambda }
\le
\frac{ 1 }{ {\rm e}^{2\kappa d} - 1 }
\label{eq:D-limits}
\end{equation}
with the stronger inequalities $0 \le 1/D_{\lambda} \le 
1/({\rm e}^{2\kappa d} - 1)$
holding for identical 
plates. In the latter case, the Casimir force is hence necessarily 
attractive. The inequalities (\ref{eq:D-limits})
saturate 
for a perfectly conducting mirror facing a perfectly permeable one
($\varepsilon_{1} = \infty$, $\mu_{2} = \infty$, say), and
for identical, perfectly reflecting mirrors, respectively.
The resulting forces at zero temperature 
are~\cite{Lifshitz56,Boyer74}
\begin{equation}
T = 0: \quad
- \frac{7}{8} F_C \le
F_L 
\le
F_C.
\label{eq:F-limits-zero-T}
\end{equation}
In the high-temperature limit, we get similarly \cite{Tort99}
$
-\frac{3}{4} F_T \le
F_L
\le
F_T \equiv \zeta(3) k_B T / (8\pi d^3)$ by keeping in
Eq.(\ref{eq:Lifshitz-formula}) only the $n=0$ term in the sum.

Consider now a mirror made from layers of arbitrary passive materials.
Reflection coefficients for such a system can be obtained recursively.
For a layer `$b$' separating a medium `$a$' from a substrate `$c$',
for example,
\begin{equation}
    r_{abc} = \frac{ r_{ab} + r_{bc} \,{\rm e}^{ 2 {\rm i} k_{b} w } 
    }{ 
    1 + r_{ab} r_{bc} \,{\rm e}^{ 2 {\rm i} k_{b} w } }
    \label{eq:multiple-layer-r}
\end{equation}
where $r_{ab}$ ($r_{bc}$) describes the reflection from the interface 
$ab$ ($bc$),
respectively, and $w$ is the layer thickness~\cite{BornWolf,Yeh}. 
If the substrate $c$ is a multilayer system itself, $r_{bc}$ is the 
corresponding reflection coefficient.
For the 
imaginary frequencies occurring in the Lifshitz 
expression~(\ref{eq:Lifshitz-formula}), the wavevector in the layer 
is purely imaginary, $k_{b} = {\rm i} \, \kappa_b$, and
single-interface coefficients are real [Eq.(\ref{eq:Fresnel-r})].
From Eq.(\ref{eq:multiple-layer-r}), they remain real for 
multilayer mirrors. In addition, the mapping $r_{ab} 
\mapsto r_{abc}$ 
is a conformal one, and if $r_{bc}
\,{\rm e}^{ - 2 \kappa_b w }$ is real and $\in[-1,1]$, 
the interval $[-1,1]$ is mapped onto itself.
For multilayer mirrors, we thus obtain again the inequalities
$-1 \le r_{\lambda} \le 1$. This generalizes the limits of
Refs.~\cite{Lambrecht97,Genet03c} that are obtained only for
layered dielectric mirrors, using transfer matrices.

\section{Casimir interaction between metamaterials}

To illustrate these generally valid results, we focus on
meta materials described by effective medium theory
\cite{Ramakrishna05,Pendry99d,Smith04a}. We 
adopt Lorentz-Drude formulas for $\varepsilon$ and $\mu$
\begin{equation}
\varepsilon_{\alpha}( {\rm i}\,\xi ) = 1 + 
\frac{ \Omega_{\alpha}^2 }{ \omega_{\alpha}^2 + \xi^2},
\qquad
\mu_{\alpha}( {\rm i}\,\xi ) = 1 + 
\frac{ \Theta_{\alpha}^2 }{ \vartheta_{\alpha}^2 + \xi^2 }
.
\label{eq:Lorentz-Drude-1}
\end{equation}
Regarding the permeability, we have taken the limit of weak
absorption and computed $\mu( {\rm i}\,\xi )$ in terms of 
${\rm Im}\,\mu( \omega )$ using the Kramers-Kronig relations.
This is necessary to ensure high-frequency transparency of
the medium. We denote in the following by $\Omega$ a
typical resonance or plasma frequency occurring in 
Eqs.(\ref{eq:Lorentz-Drude-1}). The corresponding wavelength,
$\Lambda = 2\pi c / \Omega$, provides a convenient distance
scale. Note that a (magnetic) resonance wavelength as short as 
$\sim 3\,\mu{\rm m}$ has already been achieved with
material nanofabrication \cite{Wegener04}.
The key advantage of meta materials is that their  
electric and magnetic `plasma frequencies' $\Omega_\alpha$
and $\Theta_\alpha$ are fairly large as well: a value of $\Theta_\alpha
\approx \vartheta_\alpha \sqrt{f} \le (c/a) \sqrt{f}$ is typical
for a split-ring resonator array with period $a$ and
filling factor $f$ \cite{Pendry99d}. This property is also necessary,
of course, to achieve a left-handed medium 
($\varepsilon_\alpha( \omega )$, $\mu_\alpha( \omega ) < 0$ 
for some real frequencies). The magnetic plasma
frequencies occurring in conventional ferromagnets are
much smaller~\cite{Camley98}, and the impact on the Casimir
interaction is weak, as reported recently \cite{Tomas05}.

\begin{figure}[bht]
    \resizebox{!}{58mm}{\includegraphics*{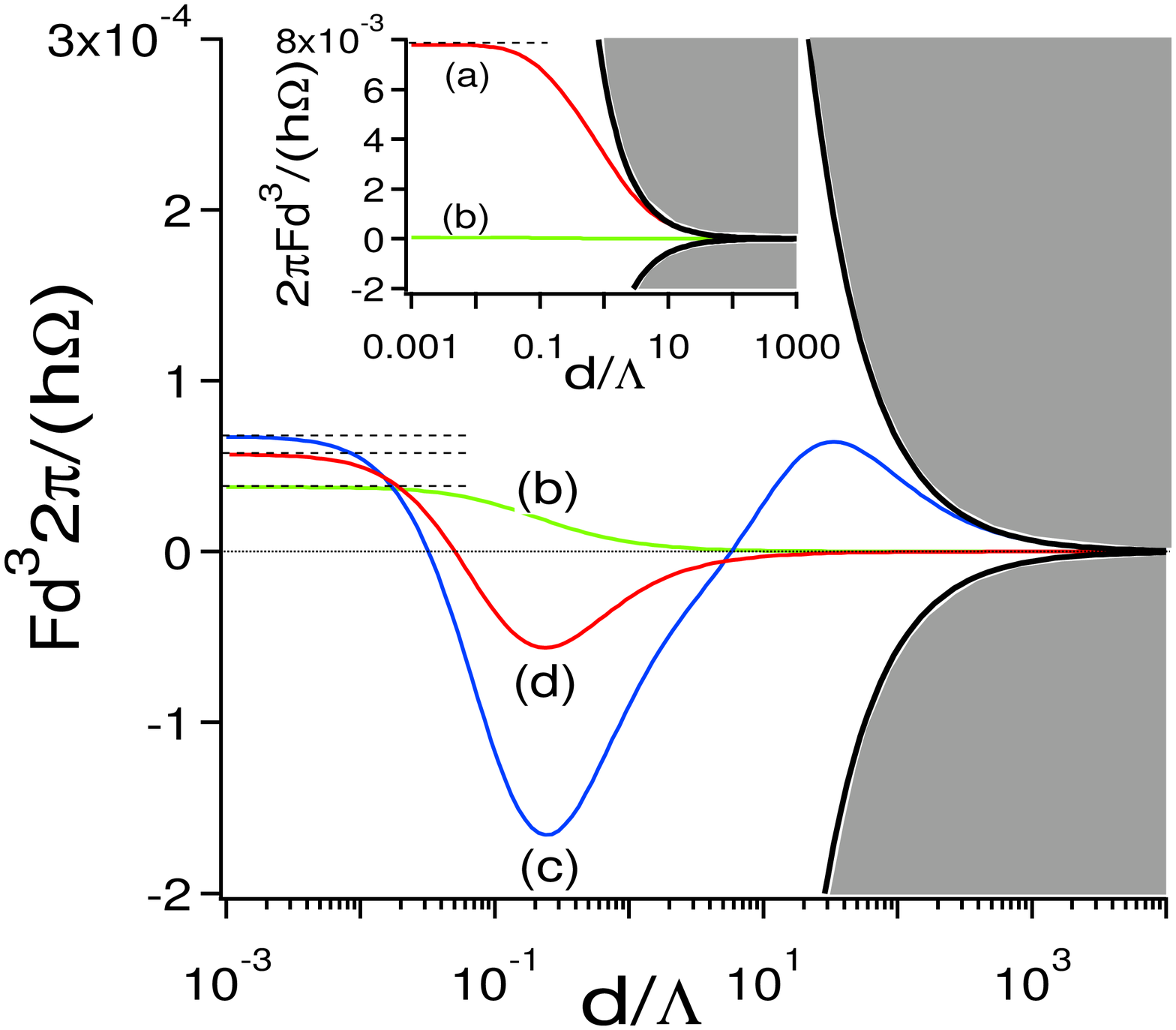}}
\hspace*{01mm}
    \resizebox{!}{58mm}{\includegraphics*{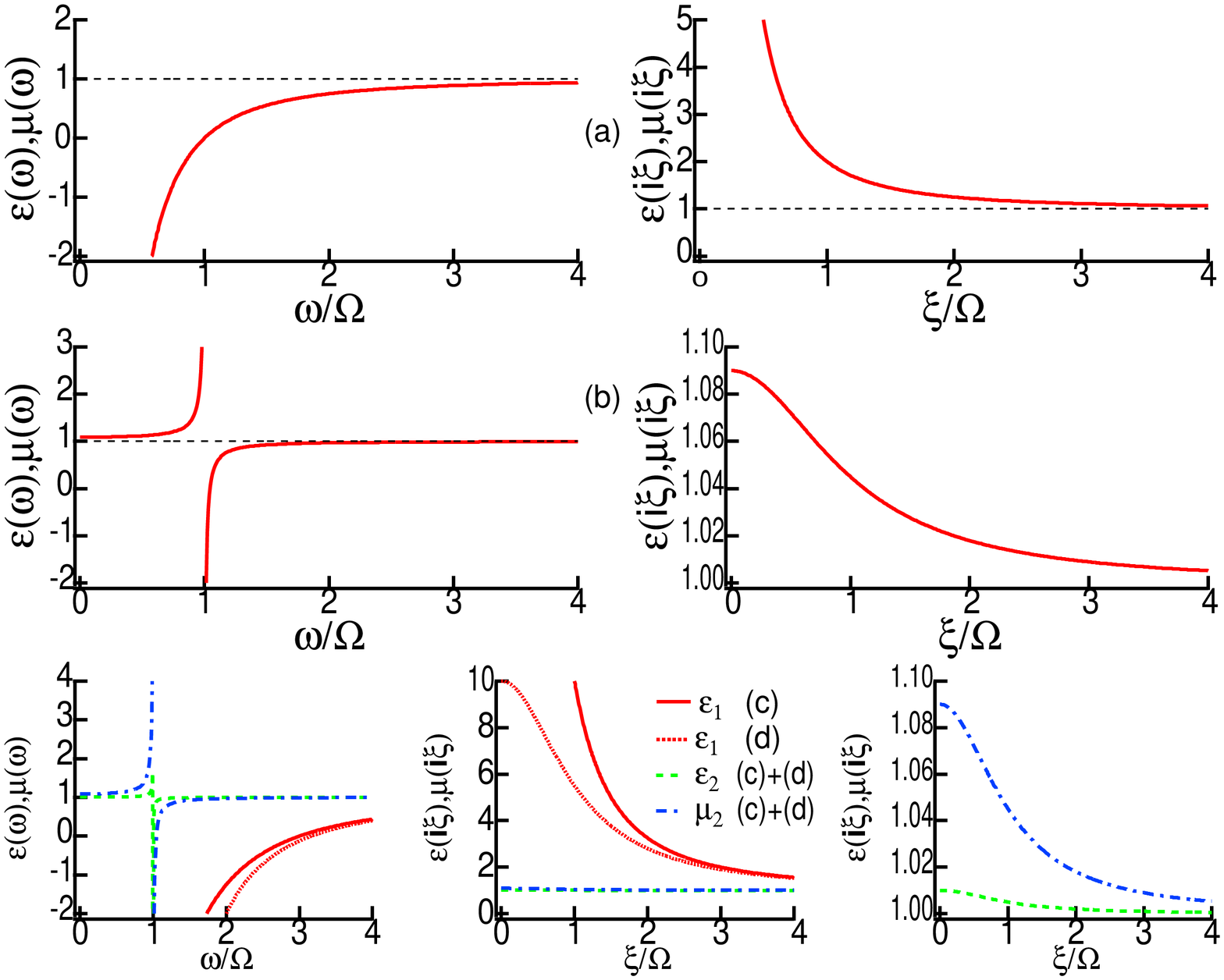}}
    \caption[]{Casimir force between planar mirrors of different
    dispersive materials at zero temperature. 
Left: plot of the force per unit area, normalized to
$\hbar \Omega / d^3$ vs.\ the plate separation scaled to
$\Lambda \equiv 2\pi c / \Omega$. Here, $\Omega$ is a 
typical plasma or resonance frequency. 
Shaded areas: excluded by the 
    inequalities~(\ref{eq:F-limits-zero-T}). 
Right: material response functions along the real and imaginary
frequency axis. 
\\
    Dashed lines: 
short-distance
	asymptotics $c_{3} / d^3$ with coefficient~(\ref{eq:c3-attraction}),
providing another upper limit for homogeneous mirrors.
    Curve (a) (inset): two identical non-magnetic Drude metals 
    (in Eq.(\ref{eq:Lorentz-Drude-1}),
    $\Omega_{\alpha} \equiv \Omega$, $\omega_{\alpha} = 0$,
    $\mu_{\alpha}( \xi ) \equiv 1$, $\alpha = 1,2$). 
    Curve (b): two identical left-handed meta materials with 
    overlapping dielectric and magnetic resonances
    ($\Omega_{\alpha} \equiv 0.3\,\Omega$, 
    $\omega_{\alpha} = \Omega$,
    $\Theta_{\alpha} =  0.3\,\Omega$,
    $\vartheta_{\alpha} = \Omega$). 
    Curve (c): two identical non-magnetic Drude metals one of which is
    coated with a left-handed meta material
    (Drude metals with $\Omega_{1} = 3\,\Omega$, $\omega_{1} = 0$,
    $\mu_{1}( \xi ) \equiv 1$;
    left-handed coating with thickness 
    $w = 10\times 2\pi c / \Omega$ and
    material parameters (dominantly magnetic response)
    $\Omega_{2} \equiv 0.1\,\Omega$,
    $\omega_{2} = \Omega$,
    $\Theta_{2} =  0.3\,\Omega$,
    $\vartheta_{2} = \Omega$).
    Curve (d): two meta materials, one purely dielectric, the other mainly 
    magnetic 
    ($\Omega_{1} = 3\,\Omega$, 
    $\omega_{1} = \Omega$, $\mu_{1}( \xi ) \equiv 1$,
    $\Omega_{2} = 0.1\,\Omega$, $\omega_{2} = \Omega$,
    $\Theta_{2} =  0.3\,\Omega$,
    $\vartheta_{2} = \Omega$). 
}
\label{fig:attr-rep}
\end{figure}

\begin{figure}[bth]
    \resizebox{70mm}{!}{\includegraphics*{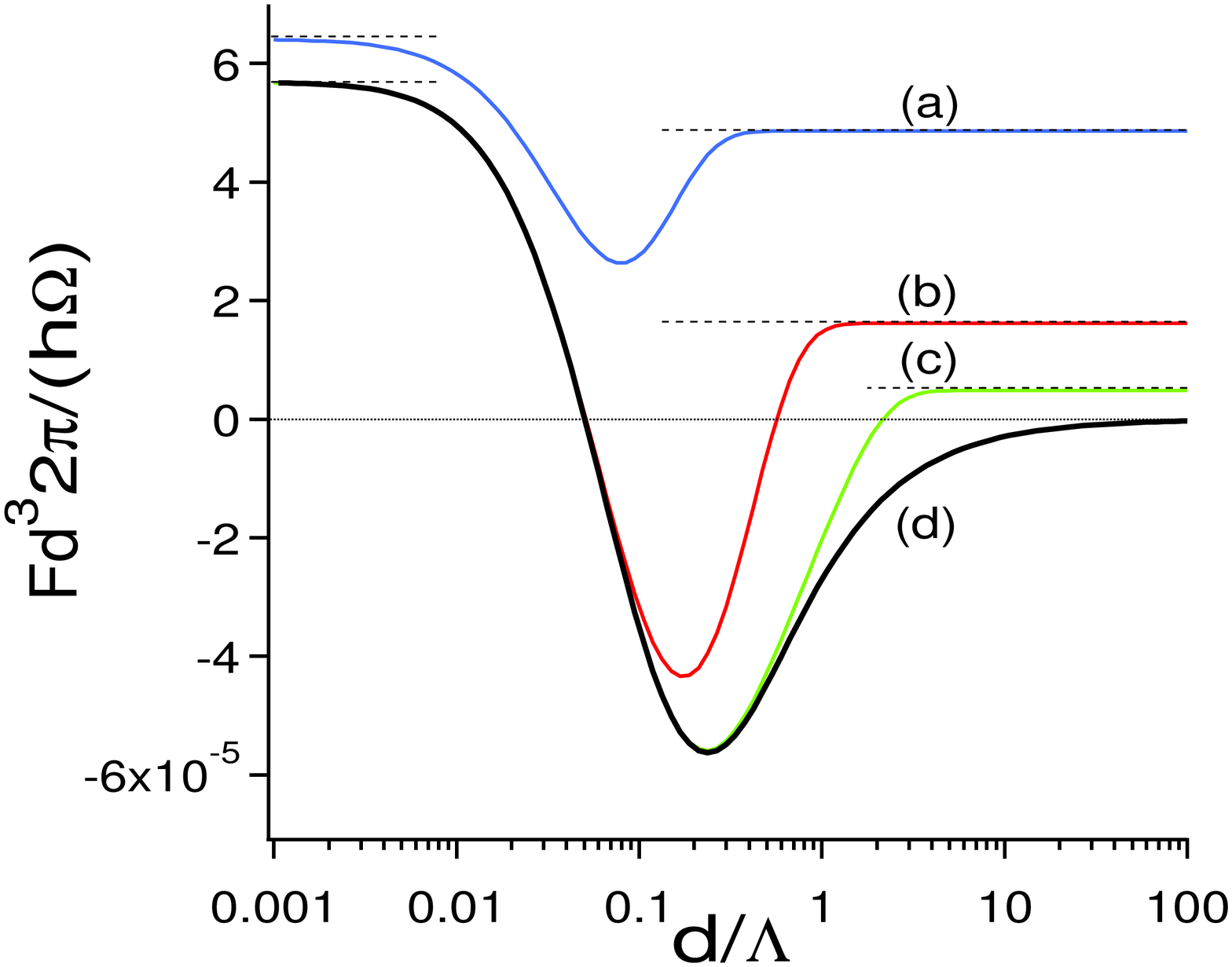}}
    \caption[]{Casimir force between two different meta materials 
    as the temperature is raised. The material parameters are 
    identical to Line (d) of Fig.\ref{fig:attr-rep}. The temperature
    takes the values $k_{B} T = (a) 0.3, (b) 0.1, (c) 0.03, (d) 0  
    \,\hbar\Omega$.
    Same scaling as in Fig.\ref{fig:attr-rep}. Dashed lines correspond 
    to the short- and long-distance asymptotics discussed in the text.
}
\label{fig:finite-T}
\end{figure}

In the plots shown below, the Casimir pressure is normalized 
to $\hbar \Omega / d^3$ [see after Eq.(\ref{eq:c3-attraction})].
In order of magnitude, this corresponds to 
$10^4\, {\rm pN}\, {\rm mm}^{-2}/(\Lambda / \mu{\rm m})^4$ 
at a distance $d = \Lambda/2$.
This can be measured with sensitive torsion balances 
\cite{Lamoreaux97a,Chan01} or cantilevers \cite{Mohideen98,Onofrio02}.
We plot in Fig.\ref{fig:attr-rep} the result of
a numerical integration of Eq.(\ref{eq:Lifshitz-formula}),
the curves corresponding to different material pairings.
One sees that in all cases, the force satisfies the
limits~(\ref{eq:F-limits-zero-T}) that exclude the shaded areas.
We observe that materials 
with negative index of refraction around $\Omega$ show a strongly
reduced attraction (Fig.\ref{fig:attr-rep}(b)). This can be attributed to
the reduced mirror reflectivity due to impedance matching. 
Casimir repulsion is achieved for some distances between mirrors
made from different materials (Fig.\ref{fig:attr-rep}(c,d)).
At short distance, i.e. $d \ll \Lambda / 2\pi$, even these pairings show
attraction with a power law $1/d^3$.
Coating one mirror with a magnetic layer (Fig.\ref{fig:attr-rep}(c)), 
there is a sign change around the layer 
thickness $w$: for $\Lambda/2\pi \ll d \ll w$, 
the layer behaves like a thick plate,
and its material parameters lead to repulsion.
The layer can be ignored for $w \ll d$, and one recovers the 
attraction between the (identical) substrates. This is consistent with 
asymptotic analysis based on
the reflection coefficient~(\ref{eq:multiple-layer-r}), as we outline
below.
Detailed calculations show that a large resonance frequency is
not sufficient to achieve repulsion, the oscillator strength of the 
resonances (proportional to $\Omega_{1}$ and $\Theta_{2}$) must be 
large enough so that $\hbar \Omega_{1}, \hbar \Theta_{2} \gg 
\max( k_{B} T, \hbar c / d )$.
As the temperature is raised, the distance range where
repulsion is observed disappears, see Fig.\ref{fig:finite-T}.
One then finds a $1/d^3$ power law at large distance as well.

The different regimes of Fig.\ref{fig:attr-rep} can be understood from
an asymptotic analysis of Eq.(\ref{eq:Lifshitz-formula}).
At short distance ($d \ll \Lambda/2\pi$), the integral is 
dominated by a region in the $\kappa$-$\xi$-plane 
where the Fresnel
coefficients (\ref{eq:Fresnel-r}) take the nonretarded forms $r_{{\rm TM}}
    \to R( \varepsilon) \equiv 
    (\varepsilon - 1)/( \varepsilon + 1 ) > 0$
assuming that $\varepsilon > 1$ and similarly $r_{{\rm TE}} \to R( \mu ) 
> 0$ unless $\mu = 1$.
Proceeding like Lifshitz \cite{Lifshitz56,Henkel04a},
yields to leading order a power law $F_{L} = c_{3} / d^3$ with a
\emph{positive} Hamaker constant given by
\begin{equation}
c_{3} = \frac{ k_{B} T }{ 4\pi }
    {\kern 4.5ex\raisebox{0.25ex}{$'$}\kern -4.5ex}\sum\limits_{n=0}^{\infty} 
\left\{
{\rm Li}_{3}[
R( \varepsilon_1(\xi_n) )
R( \varepsilon_2(\xi_n) )
] 
+
{\rm Li}_{3}[
R( \mu_1(\xi_n) )
R( \mu_2(\xi_n) )
]
\right\}
,
\label{eq:c3-attraction}
\end{equation}
where
${\rm Li}_{n}( z ) \equiv \sum_{k=1}^\infty z^k / k^n$. 
It must be noted that for the special case of homogeneous plates,
this asymptotic expression actually 
provides another, much stricter,
upper limit to the Casimir force, since
$r_{{\rm TM}\alpha} \le R( \varepsilon_{\alpha} )$, $r_{{\rm
TE}\alpha} \le R( \mu_{\alpha} )$ and ${\rm Li}_{3}( z )$ is a
monotonous function (see Fig.\ref{fig:attr-rep}).
In order of magnitude, $c_{3} \sim \hbar\Omega$ 
at low temperatures ($k_{B} T \ll \hbar \Omega$). 
Compared to ideal mirrors, 
dispersive plates thus show a much weaker Casimir interaction 
that is in general attractive (Fig.\ref{fig:attr-rep} and 
Ref.\cite{Lambrecht97}).
At larger distances, 
$\Lambda/2\pi \ll d \ll \Lambda_{T} \equiv \hbar c / k_{B} T$, 
the Casimir force follows a $1/d^4$ power law, and 
repulsion is found provided one of 
the materials is dominantly magnetic. Here, the 
non-dispersive results of Ref.\cite{Kenneth02} 
are recovered. 
Finally, for $d \gg \Lambda_{T}$,
the leading order force is the term $n=0$ in the sum 
(\ref{eq:Lifshitz-formula}),
again an attractive $1/d^3$ law, with a 
coefficient given by 
an expression similar to~(\ref{eq:c3-attraction}), but
involving the static material constants, 
see~\cite{Kenneth02}. 

The impact of temperature is illustrated in Fig.\ref{fig:finite-T}:
at high temperature, $k_{B}T \gg \hbar\Omega$, the second Matsubara 
frequency $\xi_{1}$ falls already into the mirrors' transparency zone,
and the $1/d^3$ power law is valid at all distances. 
As $T \to 0$, the intermediate repulsive zone appears in the
range $\Lambda/2\pi \ll d \ll \Lambda_T/2\pi$.
A good agreement with the analytical $1/d^3$ asymptotics is 
found outside this zone, as shown by the dashed lines.
For the resonance wavelength $\Lambda = 3\,\mu{\rm m}$ mentioned
above, cooling to a temperature $T \approx 0.1 \,\hbar\Omega/ k_B
\sim 50\,{\rm K}$ is required to `open up' the repulsive window.
This temperature increases, of course, with materials whose
response extends to higher frequencies. 

\begin{figure}[bth]
    \resizebox{70mm}{!}{\includegraphics*{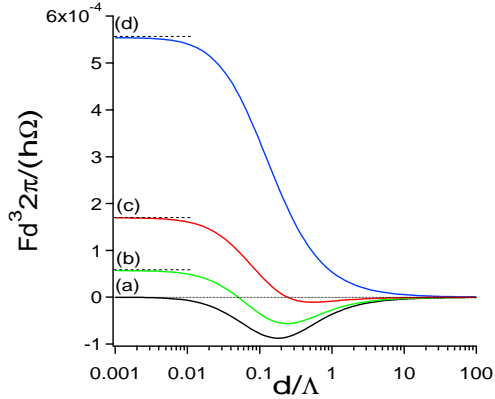}}
    \caption[]{Casimir force between two different meta materials  
    as the residual dielectric response of mirror~2 is changed. 
    Same scaling as in Fig.\ref{fig:attr-rep}, $T = 0$. 
    Dashed lines:
    short-distance asymptotics (\ref{eq:c3-attraction}).
The dielectric response for mirror~2 is of Lorentz-Drude 
form with fixed cutoff frequency and variable oscillator strength.
    Material functions as in Fig.\ref{fig:attr-rep}(d), 
    except
    $\varepsilon_{2}( 0 ) = 1$ (a), $1.01$ (b), $1.03$ (c), $1.1$ (d).
    }
\label{fig:mismatch}
\end{figure}

Finally, we would like to illustrate the kind of peculiar
asymptotics that becomes possible with carefully matched material
parameters. 
This follows Ref.\cite{Parsegian69} that computes the Van
der Waals force on a water film coated on both sides by lipid 
membranes,
finding a weak dependence on the ultraviolet frequency range because
both materials have a similar electron density.
Consider thus a liquid-filled gap with a similar electron density as
medium 2 so that $\varepsilon_{0} = \varepsilon_{2}$, and a 
permeability $\mu_{0} = \mu_{1} \equiv 1$ matched to medium~1. 
For simplicity, we assume that these equalities hold at
all frequencies. In this case, we can show that the force is repulsive
at all distances, even at finite temperature.
Indeed, both contributions in Eq.(\ref{eq:c3-attraction}) vanish,
and the leading order term for high temperatures also
vanishes.
The high-temperature limit is given by the $n=1$ term 
in (\ref{eq:Lifshitz-formula}). 
This gives a distance dependence proportional to 
$\exp( - 4 \pi d / \Lambda_{T})$ similar to what has been 
observed in some experiments with colloids (mentioned in \cite{Ackler96}).
As the temperature is lowered, this exponential 
regime still applies for $d \gg \Lambda_{T}$. 
If $k_{B} T \ll \hbar \Omega$, 
the $1/d^4$ regime of Ref.~\cite{Kenneth02} 
exists at intermediate distances $\Lambda/2\pi \ll d \ll 
\Lambda_{T}/2\pi$. The short distance regime sets in for
$d \ll \Lambda$, and an analysis similar to the one 
leading to Eq.(\ref{eq:c3-attraction}) gives ($T = 0$)
\begin{equation}
F_{L}  = \frac{ \hbar }{ \pi }
\sum\limits_{n=1}^{\infty}
\left( 2 n d \right)^{2n-3}
\int\limits_0^\infty\!\frac{ {\rm d}\xi }{ 2\pi }
\Gamma( 3 - 2n, 2 n \xi \, d / c ) 
\left(
-
\frac{ \varepsilon_{1} - \varepsilon_{0} }{ 
\varepsilon_{1} + \varepsilon_{0} }
\frac{ (\mu_{2} - \mu_{1}) \xi^2 }{ 4 c^2 }
\right)^n
\label{eq:c1-repulsion}
\end{equation}
with 
$\Gamma( k, z ) \equiv \int_{z}^{\infty}\!{\rm d}t \,t^{k-1} 
{\rm e}^{-t}$. At short distance, 
the sum is dominated by the
first term, so that to leading order, we get a repulsive power 
law $F_{L} = - c_{1} / d$ (Fig.\ref{fig:mismatch}(a)).
In order of magnitude, $c_{1} \sim \hbar \Omega^3 / c^2$
and therefore again $-F_{L} \ll F_{C}$,
with a cross over occurring around $\Lambda/2\pi$ (see 
Fig.\ref{fig:mismatch}(a)).
Due to our assumption of a perfect matching $\varepsilon_2 = 
\varepsilon_0$, this kind of behaviour seems quite
remote from experimental reality. As shown in 
Fig.\ref{fig:mismatch}(b--d),
a slight mismatch between the dielectric functions
of liquid and plate leads back to an attractive
force, first at short distances, then suppressing the repulsive
window altogether.

\section{Conclusion}

We have generalized strict upper and lower
limits for the Casimir force.
We have shown that a strongly modified Casimir force can occur 
between dispersive and absorbing mirrors with a sufficiently
large magnetic susceptibility, extending results restricted
to non-dispersive materials~\cite{Kenneth02}. The most promising 
way to achieve this repulsion seems
the use of meta materials engineered at scales between the nanometer and 
the micron because they provide a fairly large magnetic oscillator 
strength. Our results are intrinsically limited to distances
$d \gg a$ by our use of effective medium theory. Sufficiently
small structures and sufficiently low temperatures then ensure
that in the range $a \ll d \ll \Lambda_T/2\pi$, the Casimir interaction
can be strongly altered: even if repulsion cannot be
achieved in a first step, we expect a significant reduction of 
the Casimir attraction at distances of a few microns 
(Fig.\ref{fig:attr-rep}).

Our thanks for comments and discussion goes to 
J.-J. Greffet, J.-P. Mulet, M. Wilkens, M. Toma\v{s}, C. Genet, A. Lambrecht,
S. Reynaud, and L. Pitaevskii. We thank anonymous referees for
constructive criticism. We acknowledge financial support from
the bilateral French-German
programme ``Procope'' under project numbers 03199RH and D/0205739.

\newcommand{\bibpath}{/Users/carstenh/Biblio/Database/}
\newcommand{\bstpath}{/Users/carstenh/Biblio/Database/bst/}



\end{document}